\documentclass[preprint,aps,showpacs]{revtex4}

\usepackage{graphicx}
\usepackage{dcolumn}
\usepackage{bm}

\begin{document}

\title {Direct simulations of small multi-fermion systems}
\author{Michael Creutz}
\affiliation{
Physics Department, Brookhaven National Laboratory\\
Upton, NY 11973, USA
}
\email{creutz@bnl.gov}

\date{\today}

\begin{abstract}
I explore computer simulations of the dynamics of small
multi-fermion lattice systems.  The method is more general, but I
concentrate on Hubbard type models where the fermions hop between a
small number of connected sites.  I use the natural mapping of fermion
occupation numbers onto computer bits.  Signs from fermion interchange
are reduced to bit counting.  The technique inherently requires
computer resources growing exponentially with the system volume; so,
it restricted to modestly small systems.  Large volume results would
require combining these techniques with further approximations,
perhaps in a recursive renormalization group manner.
\end{abstract}

\maketitle

\def\eqn{}





Quantum systems involving fermions have proven elusive for computer
simulation.  A plethora of cancellations in all but a few cases
impedes the use of Monte Carlo methods, which have been so successful
for bosonic systems.  Here I explore the direct application of the
relevant Hamiltonians to wave function representations stored in
computer memory.  I work with large but sparse matrices acting in a
finite dimensional Hilbert space.  As the methods are inherently
exponential in volume, I concentrate on smaller systems.  Thus I am
admitting defeat in terms of a thermodynamic limit, hoping instead to
find interesting physics with finite systems.  To obtain results for
larger volumes, these techniques could form the basis for approximate
techniques, such as solving small blocks of variables to form the
starting point for a renormalization group approach.  Note that I am
discussing directly simulating the behavior of quantum systems on a
classical computer.  In this way one can also study various models for
quantum computers.  The storage required does grow exponentially with
the number of q-bits under study, but if this is fairly modest, say of
order 20, the methods exploited here are straightforward.

Consider a Fock basis $|n_0,\ldots,n_{N-1}\rangle$ for a many fermion
system, where the $n_i$ are the occupation numbers for some set of
orthogonal single particle states.  Each $n_i$ is either 0 or 1.  For
a lattice model these might represent the occupations on given sites
with given spins.  This basis naturally maps onto computer words,
which are sets of bits also either 0 or 1.  A 1 bit is ``set''
and a 0 bit ``unset.''  This raw mapping is instinctive for numerical
simulations of multi fermion systems.  A subroutine representing a
creation operator for a fermion would set the corresponding bit in the
appropriate word.  An annihilation operator resets the bit.  Such can
all be done with simple bitwise logical operations.  This naive
observation, however, requires embellishment so that fermion exchanges
will give rise to the appropriate relative negative signs.  Simple bit
counting techniques allow us to track these.

To start, consider $N$ creation/annihilation pairs
$\{a_i^\dagger,a_i\}$ for $0<=i<N$.  These satisfy the usual fermion
commutation relations
\begin{eqnarray}
[a_i,a_j^\dagger]_+ &=& a_ia_j^\dagger+a_j^\dagger a_i=\delta_{ij} \\
{} [ a_i, a_j ]_+ &=& 0
\end{eqnarray}

The vacuum state $|0\rangle$ is annihilated by all $a_i$; i.e. impose
$a_i|0\rangle=0$ for every $i$.  A general Fock state is given by
applying creation operators to this state
\begin{equation}
|f\rangle=|n_0,\ldots,n_{N-1}\rangle = (a_0^\dagger)^{n_0}\ldots
(a_{N-1}^\dagger)^{n_{N-1}}|0\rangle
\end{equation}
Each of the occupation numbers $n_i$ is either zero or one.  Note that
my sign conventions are buried in the ordering convention with
increasing index.  The concept being explored here identifies each of
these basis states with an integer $n(|f\rangle)$ whose binary
representation is given by the corresponding occupation numbers
\begin{equation}
n(|f\rangle)=\sum_{i=0}^{N-1} n_i 2^i 
\end{equation}
Given a Fock state, the occupation number of a particular site is then
easily determined by testing whether the bitwise logical operation
$n\&(1<<i)$ is zero or not.  (I assume the reader is familiar with
standard C notation for logical operations.)

Of course, in a computer the natural word length is finite, 32 bits in
today's typical personal computer, 64 in most larger machines.  If we
want to study systems of more fermions, we need to combine several
words into a higher precision integer.  This technicality is
straightforward, and I will not discuss it further here.

A general quantum state is a superposition of states in this Fock
basis
\begin{equation}
|\psi\rangle = \sum_n \psi_n |n>
\end{equation}
This involves specifying a complex number $\psi_n$ for each integer
representing one of our Fock states.  The computer storage required to
hold this information grows exponentially.  Indeed, if I need to keep
all $N$ bit states, I need storage for $2^N$ complex numbers.  In many
cases symmetries allow us to reduce this number considerably, although
the basic growth with $N$ remains exponential.

One particularly useful symmetry occurs when fermion number is
conserved.  Then for a given filling $N_f$, we only need to keep track
of integers up to $2^N$ containing exactly $N_f$ set bits.  For half
filling this involves a memory saving by a factor of order $\sqrt{\pi
N/2}$, while for other fillings the saving is greater.  If we deal
with two species of fermions each of which is separately conserved,
such as the Hubbard type models discussed later, the resulting savings
can be even more.  To enumerate states with a given total occupation
it is useful to have a function that returns the next integer with the
same number of bits as its argument; an implementation is discussed
briefly in Appendix A.

Another symmetry in many systems is translational invariance, which
manifests itself in momentum conservation.  For a given momentum,
states which are related by translation have their wave functions
related by a phase, and thus only one of them needs to be stored.
For my discussion here I will not make use of this symmetry.

So given $N_s \le 2^N$ states to be stored, we can do this in various
ways.  One is to use a hash table, keying wave function components to
the corresponding Fock states, as discussed in \cite{refmc}.  If the
Fock states are ordered in some manner, one can instead use a binary
search.  This is the method used here, where for each wave function
under consideration I keep an array of $N_s$ complex numbers.  By
having the state table ordered, a given state can be quickly located.
Then the corresponding location in the coefficient table contains the
desired component of the wave function.

As storage represents the main bottleneck in this type of algorithm,
tricks to reduce this are desirable.  If we can deal with a real
Hamiltonian, the wave function storage drops by a factor of two.  With
several species each separately conserved, the table of states
separates into multiple tables representing the Fock states for the
individual species.  In this case the bulk of the storage is for the
complex numbers representing the wave function.  In any case, the type
of simulation discussed here spends the dominant amount of its
computer time doing the searches through these tables for desired
states.  Floating point arithmetic operations tend to be
insignificant.

I now become more specific and consider an annihilation operator
$a_j$.  I implement this as subroutine which takes as arguments the
specific location $j$ and a pointer to a Fock state labeled as above
by an integer $n$.  It should return zero unless bit $j$ is set in
$n(|f\rangle)$.  A simple test for this is whether, the integer
$n\&(1<<j)$ is non-vanishing.  When this is true the function returns
the corresponding sign obtained after the annihilation flips this bit.
The resulting Fock state is represented by $n\wedge(1<<j)$.  The sign
returned is associated with bringing the operator $a_j$ into the
canonical ordering above.  This is determined by the parity of the
number of set bits lower than $j$.  This can be quickly found by
logical operations, beginning with a masking off of the lower bits by
the considering the integer $n\&(1-(1<<j))$.  We need to include a
negative sign if the population count of this integer is odd.  The
population count of an integer is the number of set bits it contains.
Appendix A discusses one simple way to implement such a count.  For a
creation operator, one does exactly the same thing except checking
that the initial bit is not set.

To apply some combination of creation and annihilation operators to a
wave function $|\psi\rangle$ involves looping over all the component
Fock states $|f\rangle$.  On each of these in succession we apply the
above subroutines, multiply the component of the wave function by the
returned sign, and then store the result.  For a simple hopping
Hamiltonian, it is useful to make a table of the bit locations for the
neighbors of any given site. This small array is set up once at the
beginning of a simulation.  Then the hopping term in the Hamiltonian
becomes another subroutine which loops over sites, spins, and
neighbors.  It successively applies the corresponding creation and
annihilation pairs to the components of the source wave function.  The
results accumulate in a destination wave function.

These bit manipulations do not depend in any deep way on the type of
fermion interaction used.  Here for convenience I consider a simple
Hubbard form \cite{refhubbard}.  Thus I will consider two types of
fermions distinguished by ``spin'' and have them interact by adding an
energy $U$ for each site with both spin states occupied.  Such an
interaction is also easily implemented by logical operations; on
looping over states one multiplies the coefficient of each component
of the wave function by $U$ times the number of doubly occupied sites.
The latter is the population count of $n_\uparrow \& n_\downarrow$.

To be more specific, consider a set of sites with nearest neighbors
connected by bonds.  On each site $i$ the operator $a_{i,s}^\dagger$
creates a fermion with a spin index $s\in\{\uparrow,\downarrow\}$.  My
Hamiltonian is
\begin{equation}
H=U\sum_i n_{i,\uparrow}n_{i,\downarrow}
-\sum_{\{i,j\},s} (a_{i,s}^\dagger a_{j,s}+a_{j,s}^\dagger a_{i,s})
\end{equation} 
Where $n_{i,s}=a_{i,s}^\dagger a_{i,s}$ and $\{i,j\}$ denotes the set
of neighboring pairs.  For simplicity I set the energy scale to make
the hopping parameter unity.  In one dimension this model is exactly
solvable
\cite{refliebwu}.
For a more detailed discussion of this solution see 
\cite{refliebwua}.
Monte Carlo methods for this system are discussed in Ref.
\cite{refsugar}.
In more dimensions the Monte Carlo approach only works effectively for
the half filled case \cite{refmcmc}.

I now illustrate several simple numerical ``experiments.'' I will
first play with a six membered ring of sites, thus mimicking a benzene
molecule.  With a nearest neighbor hopping, we have the standard
undergraduate example for using linear combinations of atomic orbitals
to illustrate hybridization of the pi electrons.  When $U=0$ this has
single fermion states with energies $-2,-1,-1,1,1,2$.  Filling the
lowest levels with three spin up and three spin down electrons gives a
total ground state energy of -8.  This is the energy gained from the
delocalization of the electron wave functions.  This should be
compared with the value -6 which would be obtained from three fixed
double bonds and no hybridization.

Turning on the Hubbard interaction raises the ground state energy.  As
$U$ goes to infinity the ground state energy rises, with the largest
components of the wave function alternating spin up and down around
the ring.  They prefer to alternate rather than some other pattern
since as long as $U$ is not infinity, this state maximizes
delocalization.  For this case of half filling, i.e. 3 electrons of
each spin, there are 400 Fock states.  This corresponds to twenty
possible arrangements for each set of spins.  This is not a
particularly large matrix and could presumably be treated by
conventional matrix methods, but it enables fast experiments with the
table manipulation ideas discussed here.  The computer time for these
experiments is insignificant, easily practical on a PC.

For most of the following experiments I start with a random initial
state.  To construct such, I set each of the components of the wave
function to a Gaussian random number and then normalize the state.
From this I repeatedly apply the Hamiltonian in various ways discussed
below.  Formally I work with what is known as a Krylov space.  Note
that when there are degenerate states, a Krylov procedure does not
span the full space, but leaves the relative contribution of the
degenerate states unchanged.  To separate them one must consider
multiple starting states.  For example, one can first extract a state
of interest, construct a random state orthogonal to it, and construct
a new Krylov space which will involve different combinations of the
degenerate states.  If the degeneracy is due to some symmetry, an
alternative is to use starting states which are eigenvectors of this
symmetry.

\begin{figure*}
\includegraphics{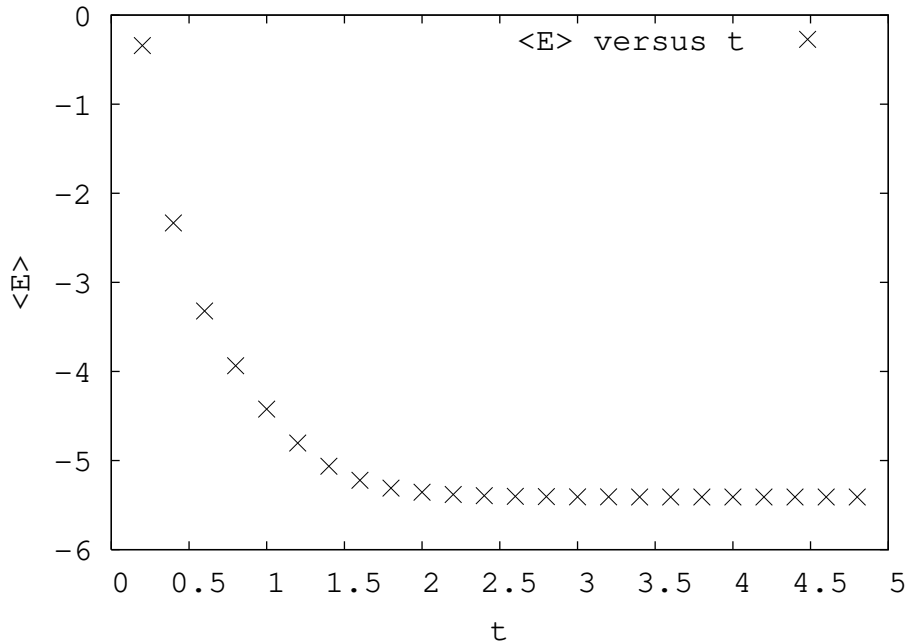}
\caption{\label{fig:emth}
The expectation value of the energy in the state
$e^{-Ht}|\psi\rangle$ with the initial $|\psi\rangle$ chosen randomly.
The system is a six membered ring with Hubbard interaction $U=2$.}
\end{figure*}

A particularly intuitive way to find the ground state is to start with
a random state as above and directly apply $e^{-Ht}$.  For large $t$
this should project out the ground state.  For moderate $t$ one can
use the rapidly convergent power series expansion for the exponential.
For larger $t$ break the evolution into smaller time intervals and
apply the exponentiated Hamiltonian repeatedly.  In Fig.~\ref{fig:emth}\ I
show the behavior of the expectation value of the energy as a function
of $t$ for our benzene system where I take the parameter $U=2$.

\begin{figure*}
{\includegraphics{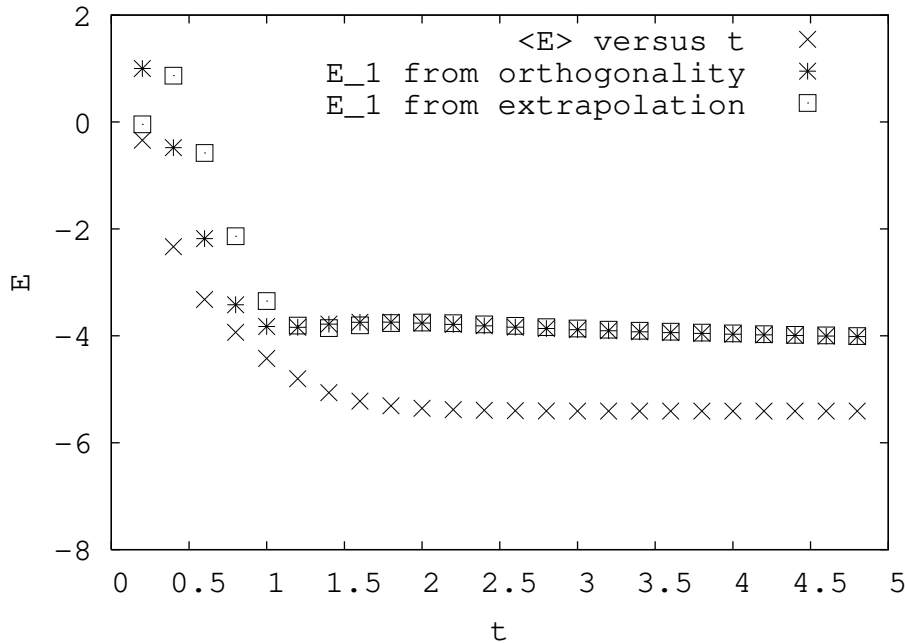}}
\caption{\label{fig:excited}
Extracting the first excited state.  Considering
$|\psi(t)\rangle=e^{-Ht}|\psi(0)\rangle$, the boxes represent the
expectation of the Hamiltonian in the combination of $|\psi\rangle$
and $H|\psi\rangle$ that is orthogonal to $|\psi\rangle$.  The bursts
represent the energy of the first exited state extracted from three
successive measurements of the expectation value of the energy.  For
comparison, the crosses replot the data from Fig.~\ref{fig:emth}.
}
\end{figure*}

There are a variety of ways to obtain information on the first excited
state from this experiment.  The value of its energy can be extracted
from the approach to the ground state via the formula
\begin{equation}
\langle E \rangle= E_0+\alpha e^{-2(E_1-E_0)t} + \ldots
\end{equation}
Solving three successive times for $E_1$ gives the results shown by
boxes in Fig.~\ref{fig:excited}.  Alternatively, as time evolves we can
extract the part of $H|\psi\rangle$ that is orthogonal to
$|\psi\rangle$ via the construction
\begin{equation}
|\psi_1\rangle=H|\psi\rangle\langle\psi|\psi\rangle
-|\psi\rangle\langle\psi|H|\psi\rangle
\end{equation}
While $|\psi\rangle$ evolves, this should be dominated by the first
excited state.  Measuring the expectation of the Hamiltonian in this
state gives the points represented by bursts in Fig.~\ref{fig:excited}.  Both
the above techniques will fail when $t$ is large enough that we have
the ground state to machine precision.

\begin{figure*}
{\includegraphics{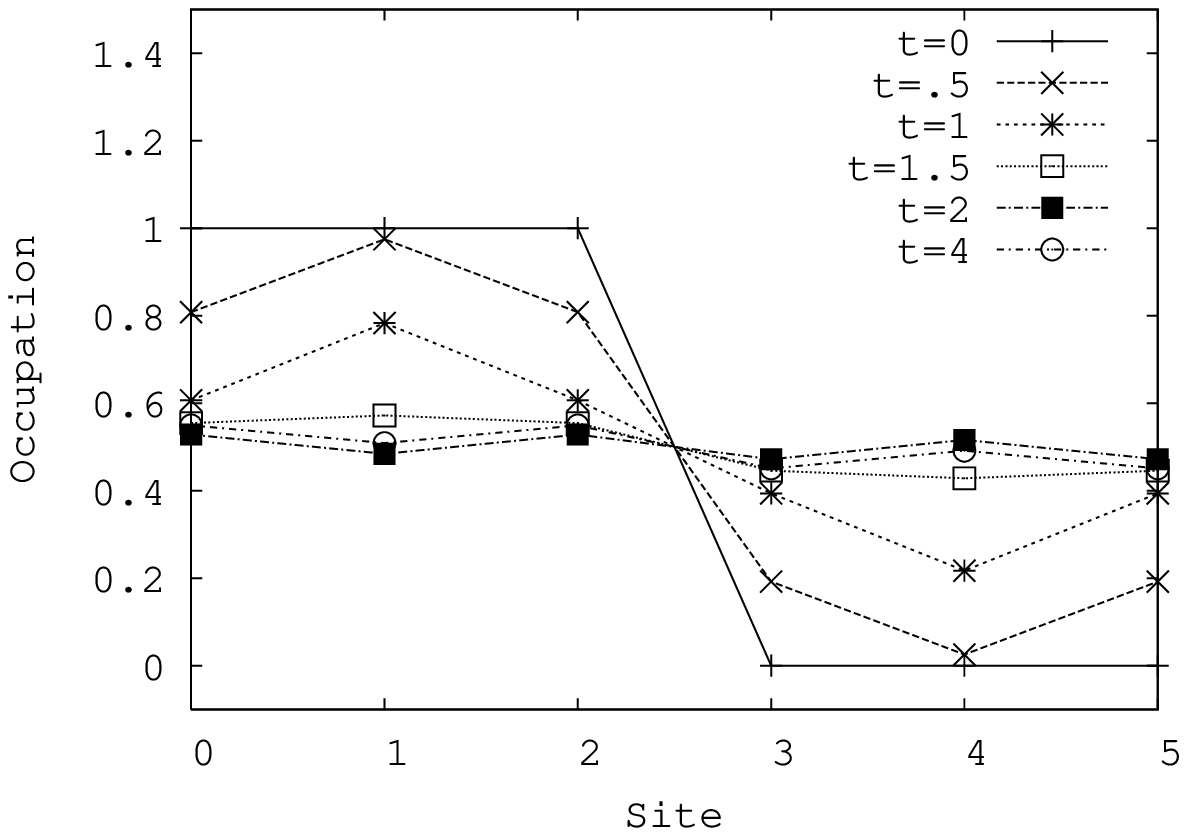}}
\caption{\label{fig:rxn} 
The relaxation of the occupation number distribution as our wave
function evolves in real time.  The initial state has all fermions on
the first three sites.  For this picture $U=2$.
}
\end{figure*}

The above evolution is effectively in ``imaginary time,'' and damps
the system to its ground state.  Since all signs are being included,
one can also work in real time and calculate the evolution of a state
under application of $e^{-iHt}$.  Since this involves no damping, it
will leave the expectation value of the energy unchanged.
Fig.~\ref{fig:rxn}\ shows a simple experiment where the initial state
has all six fermions placed on the first three sites and then observes
the expected occupation number of either spin on the sites as a
function of time.  Again I use $U=2$.  Observe the particles spreading
towards a uniform distribution.  Note that if I did not include the
interaction $U$, then all energy levels are spaced by integer amounts
and instead of a relaxation we have a sloshing of the fermions back
and forth with a periodic return to the original state.  Working in
real time provides a means to study finite temperature, or more
precisely, allows a micro-canonical evolution at energies above the
ground state.

\begin{figure*}
{\includegraphics{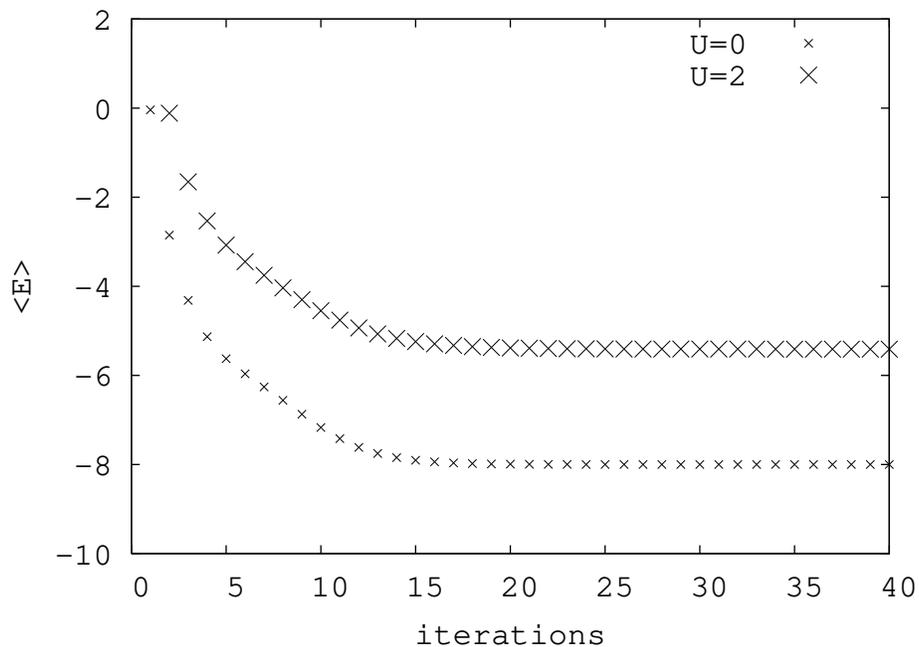}}
\caption{\label{fig:iter} The expectation value of the energy as a 
function of the number of iterations, where for each iteration I take
the linear combination the current state with the Hamiltonian on that
state that minimizes the resulting energy expectation.
}
\end{figure*}

\begin{figure*}
{\includegraphics{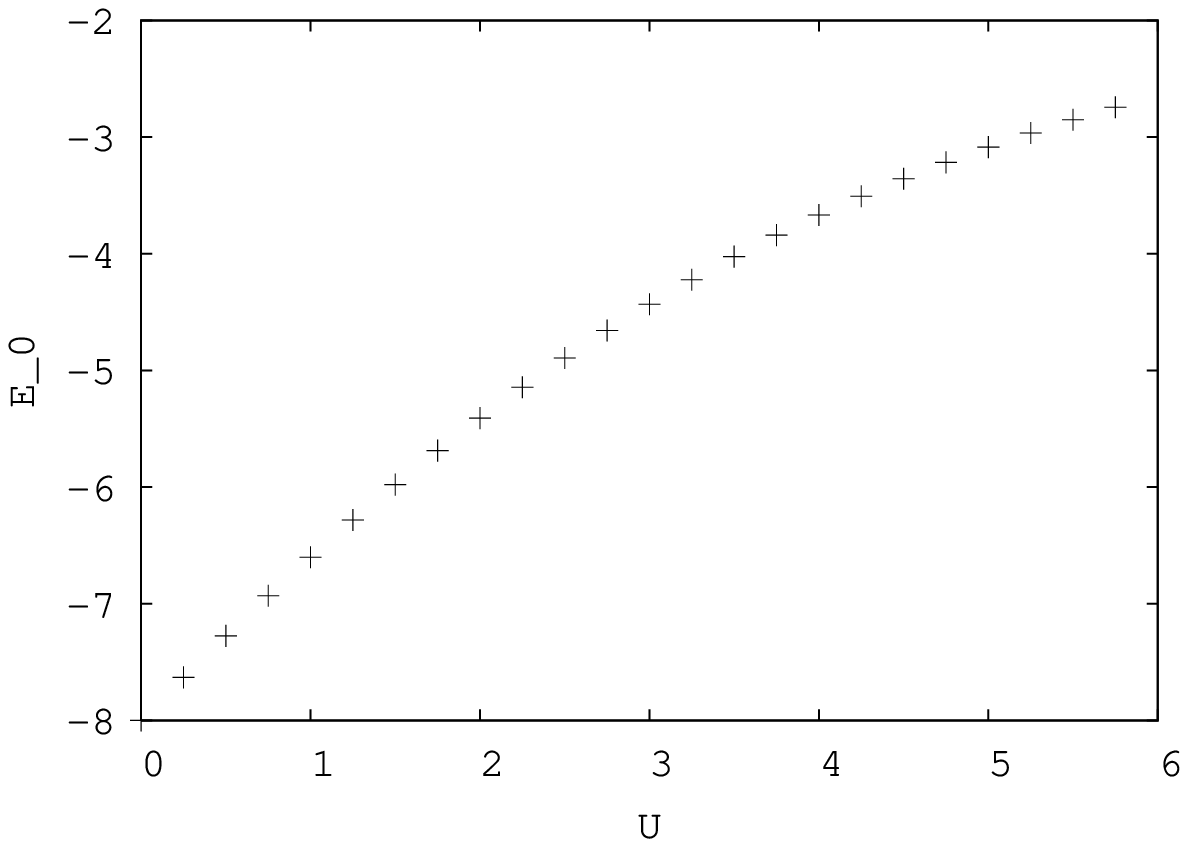}}
\caption{\label{fig:eofu} 
The ground state energy of the Hubbard model on a 6 member ring as a
function of the coupling $U$.
}
\end{figure*}

While working well for this small system, the calculation of $e^{-Ht}$
is somewhat tedious.  A more efficient but still simplistic algorithm
for finding the ground state is to repeatedly apply $H$ to the current
state and then form the linear combination of $\psi$ and $H\psi$ that
minimizes the expectation value of the energy.  At each stage this
requires calculating $H^2\psi$ as well, but is straightforward to
implement.  Fig.~\ref{fig:iter}\ shows the convergence of this
procedure for our half filled benzene system with $U=0$ and $U=2$.
Using this technique to find the ground state energy, I plot in
Fig.~\ref{fig:eofu}\ the ground state energy as a function $U$.

\begin{figure*}
{\includegraphics{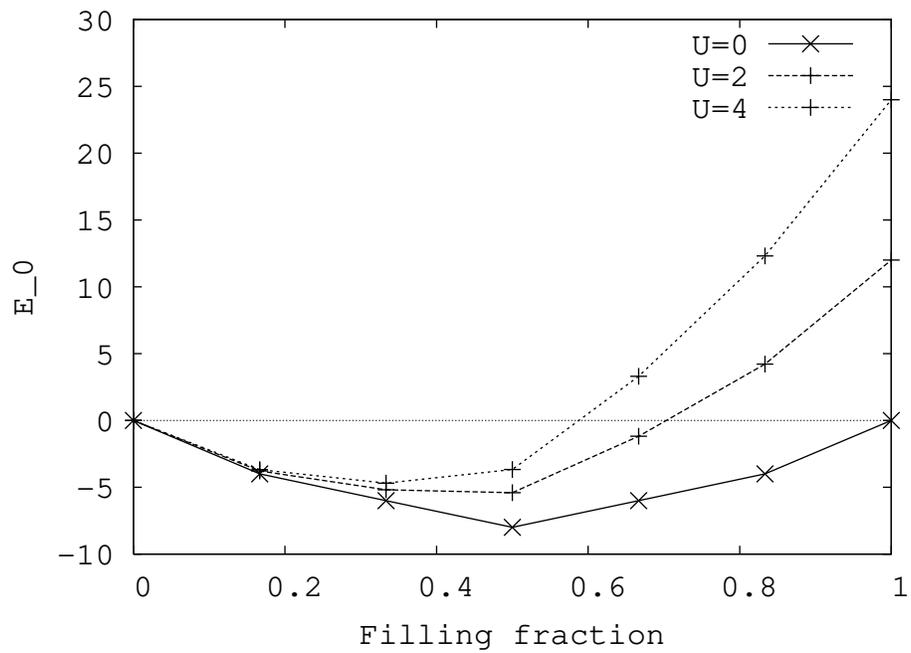}}
\caption{\label{fig:filling} 
The ground state energy of the Hubbard model on a 6 member ring as a
function of the filling for three values of the coupling $U$.
}\end{figure*}

The methods described here do not have any ``sign problems'' since all
signs are kept track of at all times.  While the above experiments
were done at half filling, there is nothing that requires this.  In
Fig~\ref{fig:filling}\ I show the ground state energy as a function of
filling fraction for the 6 membered ring.  I keep the spin up and spin
down filling fraction the same.  Note how at $U=4$ the on-site
repulsion is sufficiently strong to make the lowest energy state at
1/3 filling, rather than the 1/2 of the free case.  The $U=0$ points
in this figure show the successive fermions contribute increasing
values to the energy, demonstrating the Pauli exclusion principle as
the lower levels are filled.

The standard technique for dealing with these large sparse matrices is
the Lanczos scheme.  This iteratively constructs a sequence of vectors
$|g_i \rangle$ that form a basis under which the Hamiltonian is real
and tri-diagonal; it has non-vanishing elements only between diagonal
and sequential states in the sequence.  The construction is recursive
and makes use of an auxiliary sequence of states $|d_i \rangle$ which
satisfy a matrix orthogonality condition $\langle d_i | H |
d_j\rangle=0$ whenever $i\ne j$.  The construction
starts from an arbitrary initial $|g_0\rangle=|d_0\rangle$.  The higher
states are given by
\begin{eqnarray}
|g_n\rangle& \propto  & H|d_{n-1}\rangle - |g_{n-1}\rangle 
\langle g_{n-1}|H|d_{n-1}\rangle \\
|d_n\rangle& \propto & |g_n\rangle \langle d_{n-1}|H|d_{n-1}\rangle
- |d_{n-1}\rangle 
\langle d_{n-1}|H|g_{n-1}\rangle
\end{eqnarray}
For convenience I choose the proportionality constants so that both
vectors are normalized.  The orthogonality of the $|g_i\rangle$ and
the matrix orthogonality of the $|d_i\rangle$ are easily proven by
induction.  The matrix elements of the tri-diagonalized Hamiltonian
can be calculated during the recursion without generating any
additional vectors by expanding $|d\rangle$ 
\begin{equation}
|d_i\rangle=\sum_0^i |g_i\rangle\langle g_i| d_i \rangle
\end{equation}
and using the orthogonality constraints to obtain
\begin{eqnarray}
\langle g_{i+1}|H|g_i\rangle &=&
\frac{\langle g_{i+1}|H|d_i\rangle}{ \langle g_i|d_i\rangle }
\\
\langle g_i|H|g_i\rangle&=&
{
 {\langle g_i|H|d_i\rangle
-\langle g_i|H|g_{i-1}\rangle \langle g_{i-1}|d_i\rangle}
\over \langle g_i|d_i\rangle }
\end{eqnarray}

The iteration procedure should formally terminate at the dimension of
the Krylov space generated by applications of $H$ to $|g_0\rangle$.
If there are no degenerate eigenvalues and if $|g_0\rangle$ has
non-zero overlap with all states, then this is the dimension of our
Hilbert space.  However, any degenerate states cannot be separated in
this process, and thus the dimension of the generated space is reduced
by one for each degenerate state.  In a practical simulation on a
large system one will usually stop the series at much earlier stage.
For a small system, however, the termination is usually signaled by an
extremely large normalization factor in the construction of
$|g_n\rangle$, which would be zero were it not for finite machine
precision.  After this occurs the orthogonality with earlier $g_i$ is
lost.

\begin{figure*}
{\includegraphics{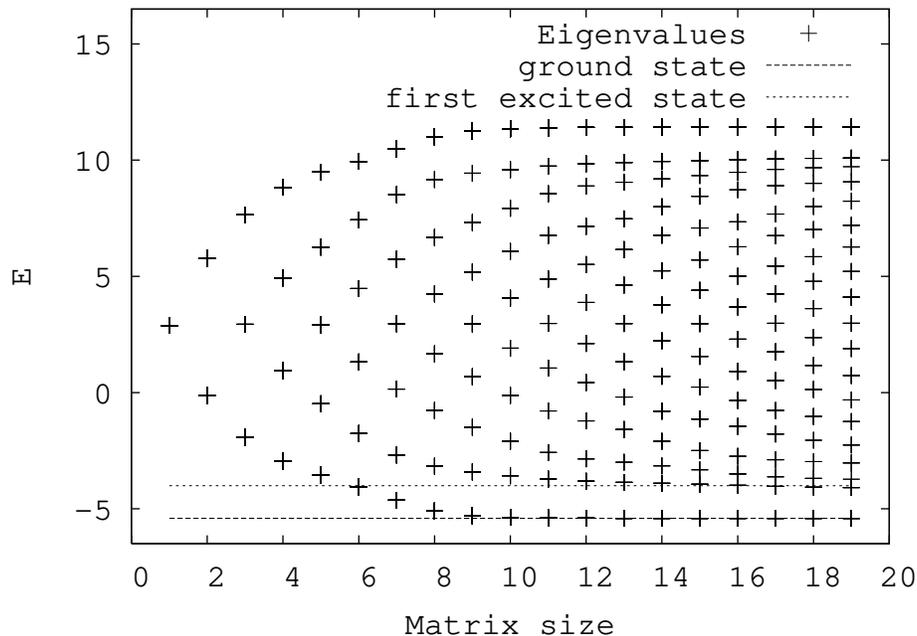}}
\caption{\label{fig:lanczos}  
The eigenvalues of the Hamiltonian in the Lanczos basis truncated to
size $n$, plotted as a function of $n$.  Note how the highest and
lowest eigenvalues stabilize the most quickly while new states appear
between them.  The horizontal lines are the ground and first excited
state energies from the last points in Fig.~\ref{fig:excited}.
}\end{figure*}

In Fig.~\ref{fig:lanczos}\ I plot the eigenvalues of the truncated
tri-diagonal Hamiltonian from our benzene system as a function of the
number of steps taken in this process.  The lowest and highest
eigenvalues rapidly converge to their respective state energies, while
new eigenvalues appear in the middle of the spectrum.  The Lanczos
procedure converges more rapidly to the true ground state energy than
the previous schemes.  For long runs, however, it can become unstable
as roundoff errors accumulate in the sequence.  The approach also
gives information on the higher levels, although, as can be seen in
the figure, intermediate states taking successively longer to
converge.  Note that for the filling discussed here both the first and
second excited states should be doubly degenerate with non-zero
angular momentum around the ring.  As discussed earlier, this single
Krylov space approach cannot separate such degeneracies.

Note how in only a few tens of iterations we have a reasonably accurate
estimate of the ground state energy.  Compared to other methods
discussed above, this involves a substantially smaller number of
applications of the Hamiltonian to our starting state, and thus is
generally regarded as the method of choice for larger systems.  Using
the states generated in this procedure as a basis reduces our 400
state system to, say, a 20 by 20 tridiagonal matrix that reproduces
well the lowest few eigenvalues.  Truncating to states of this reduced
matrix could provide a useful starting point for an approximate
iterative growth to larger systems.  Nevertheless, for the small
systems considered here, the computer time is insignificant; thus, the
earlier more intuitive but less efficient methods still work quite
well.

Fig.~\ref{fig:lanczos} displays an interesting symmetry between the
highest and the lowest energy levels.  This is a consequence of half
filling on a bipartite lattice.  Changing the sign of the fermionic
operators on half of our lattice shows that the sign of the kinetic
term does not affect the spectrum our Hamiltonian.  Doing a particle
hole transformation for a fermion $a_i\leftrightarrow a_i^\dagger$
changes $n_i \leftrightarrow 1-n_i$.  Doing this on all sites for
either value of the spin and then shifting the spectrum by the filling
of the other spin changes the sign of the potential term.  Thus,
whenever either spin state is half filled, we have a symmetry in the
spectrum under a combined shift and sign change.

\begin{figure*}
{\includegraphics{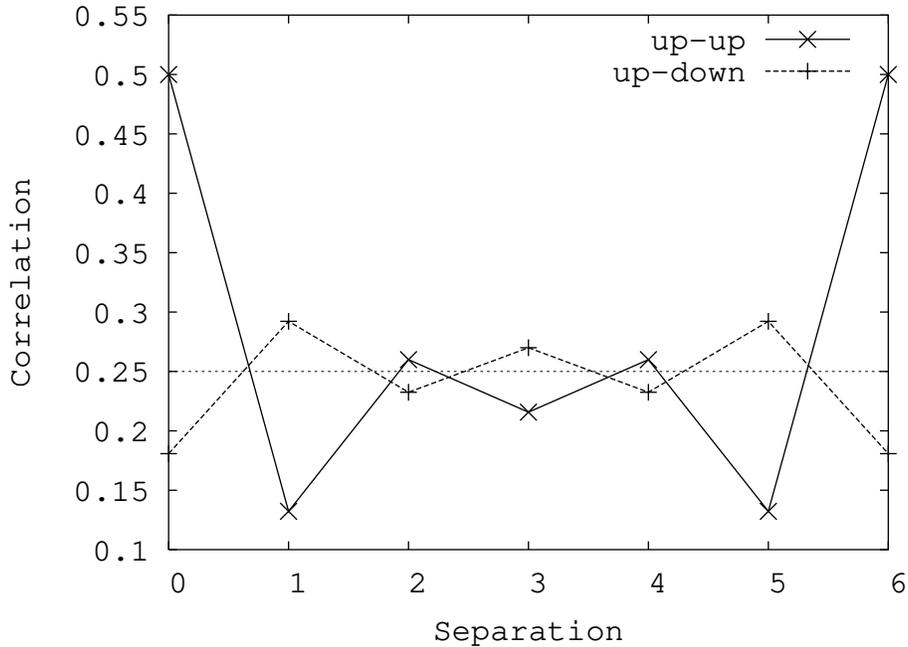}}
\caption{\label{fig:correlation}  The correlation $\langle n_{i,s}
n_{j,s^\prime}\rangle$ between spins as a function of their spatial
separation.  This is for $U=2$ on our six site cyclic system.  Note
the anti-ferromagnetic tendency.
}\end{figure*}

Having the ground state at hand enables one to look at correlations.
In Fig.~\ref{fig:correlation}\ I plot the correlation between a spin up on
one site and either spin up or spin down on another site as a function
of the separation between them.  This graph is for $U=2$ Note the
tendency for the system to become anti-ferromagnetic; such a
configuration maximizes the stabilization by delocalization.  Note
also the stronger correlation between same versus opposite spins.
When $U=0$ the up and down spins are totally decorrelated, while the
Pauli principle leaves a correlation between parallel spins.  Of
course the correlation between parallel spins on the same site is the
filling factor, .5 in this case.  For larger systems this figure
should match onto the Monte Carlo results shown in Fig.~8 of 
Ref.~\cite{refsugar}.

\begin{figure*}
{\includegraphics{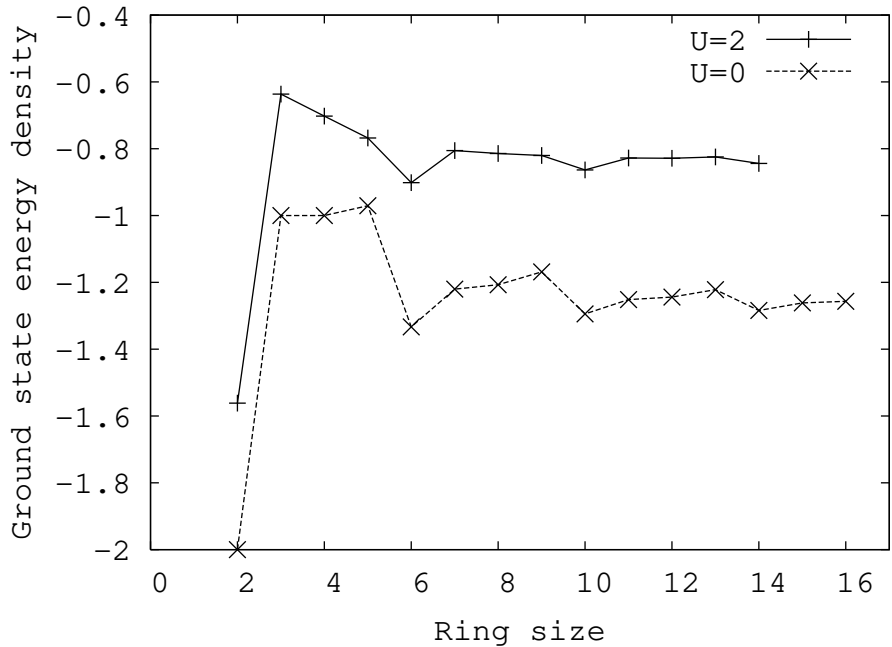}}
\caption{\label{fig:size}  
The ground state energy density $E_0/N$ as a function of the size of
an $N$ membered ring.  The upper points are for $U=2$ and the lower
ones for $U=0$.  When the interaction is turned off these numbers are
from the analytic formula; the interacting points are from the
simulation.  }\end{figure*}

The primary difficulty with these direct approaches is that memory
needs grow exponentially with system size.  Generalizing the benzene
system to an $N$ site ring, the number of basis states for the half
filled case is $\left(N!\over(N/2)!\right)^2$.  The 400 states needed
for the 6 site case rapidly rises to 853,776 for a 12 member ring.
The half filled Hubbard model on a two dimensional lattice of size 4
by 4 involves 165,636,900 states.  After a few megabytes of storage,
one leaves the realm of current personal computers.  In
Fig.~\ref{fig:size}\ I show the ground state energy density $E_0/N$ as
a function of the ring size $N$.  The filling is one half in all
cases, with the number of up and down spins differing by zero (one)
for the odd (even) rings.  The points for $U=0$ were obtained from the
analytic formula.  Note the extra stability when the ring size is
twice an odd number.  In this case Ref.~\cite{refliebwua} proved that
the ground state is unique.  For odd ring sizes the ground state
should be doubly degenerate.  This is the case in the non-interacting
case where the final fermion has non-zero angular momentum around the
ring.

I note that as with the algorithm discussed in Ref.~\cite{refmc}, this
approach does parallelize quite well if storage and computation can be
done in parallel.  If a creation or annihilation operator flips a high
bit in a Fock state, this will relate components that are far apart in
storage.  Thus the algorithm requires long range communications.
Nevertheless, the loops over components do not need the results
immediately.  Thus the results for the new wave function can be sent
off to storage while a given processor continues to work on further
components that are locally stored.  These vectors tend to be quite
long, and thus most communication is completed before the new results
are needed.  Thus we expect good performance from massively parallel
MIMD machines, including ones designed primarily for local
communication, such as the QCDSP \cite{qcdsp} and the QCDOC
\cite{qcdoc}.  As the problem is primarily combinatorial, the
performance is not determined or properly measured in terms of
floating point operations.

\appendix
\section{Counting bits}

Counting the set bits in a given computer word lies at the heart of
the discussions in this paper.  There are a variety of ways to
accomplish this.  Some computers have an assembly instruction that
directly returns this ``population count.''  However in the interest
of portability it is useful to implement this counting in a higher
level language.  A variety of fast schemes exist, but one well known
approach increments a counter while $i$ is non vanishing and
repeatedly takes $i$ to $i\&(i-1)$.  This latter operation resets the
lowest non-vanishing bit of $i$, and the repetition stops when all
bits are cleared.  Thus the following implementation in C:
\medskip
\begin{verbatim}
inline int bitcount(int i) { 
  /* counts the set bits in a word */
  int result = 0;
  while (i) {
    result++;
    i &= (i - 1);               /* finds and resets rightmost set bit */
  }
  return result;
}
\end{verbatim}
\medskip

Another useful operation given an integer $i$ is to find the next
integer with the same number of set bits.  For this first locate the
lowest run of set bits.  Move the highest of these up one position and
slide the remainder down to start at bit 0.  For example, if our
integer in binary is, say, (0011001110), we take the run of three set
bits from position 1 to 3, move the highest of these to position 4,
drop the other two to the beginning of the word, and obtain
(0011010011) as the next integer with five set bits.  This leads to
the following implementation
\medskip
\begin{verbatim}
int nextone(int i){
  /* find the next integer with the same bitcount as i */
  int bit=1,count=-1;
  if (i==0) return 1<<nsites;
  /* find first one bit */
  while (!(bit&i)){
    bit<<=1;
  }
  /* find next zero bit */
  while (bit&i){
    count++;
    bit<<=1;
  }
  if (!bit) die("overflow in nextone");
  i &= (~(bit-1));           /* clear lower bits */
  i |= bit | ((1<<count)-1); /* put them in new places */
  return i;
}
\end{verbatim}
\medskip
Defining a variety of routines for manipulating wave functions can be
quite useful.  For these I define a ``wavefunction'' type as a pointer
to a complex array.  Once a generic set of routines is set up, one can
quickly run through a variety of experiments as discussed above.  Some
such functions whose action should be clear from their names are
\begin{verbatim}
double complex overlap(wavefunction psi1, wavefunction psi2); 
double norm2(wavefunction psi); 
double normalize(wavefunction psi); 
void cmultiply(double complex factor, wavefunction psi); 
void caxpby(double complex a, wavefunction dest,
\end{verbatim}

\end{document}